# Artificial Human Phantoms: Human proxy in testing microwave apparatus that have electromagnetic interaction with the human body

A. T. Mobashsher, A. M. Abbosh

School of ITEE, The University of Queensland
St. Lucia, Brisbane QLD 4072, Australia
E-mail: a.mobashsher@uq.edu.au, a.abbosh@uq.edu.au

## I. Introduction

Quantitative and explicit validation of the performance and safety of microwave systems and devices that have electromagnetic interaction with the human body are critical factors of the technological development process. Although a numerical model of the system environment can be ideally simulated, it cannot reflect the realistic environment which is vulnerable to various electrical, mechanical and environmental interferences. Hence, the presence of human body is the best measurement environment for these systems. However, newly designed devices/systems that rely on the human body-electromagnetic fields interactions require multiple tests/measurements under a controlled environment. That environment is needed to validate the performance at all the possible scenarios of operation, and make sure of the safety of those devices and systems. For example, a breast imaging system is needed to be evaluated by detecting tumors in multiple locations and it is unthinkable to do that on a real patient; hence breast phantoms are demanded in order to obtain optimum system design and algorithms before moving to human clinical trials. Moreover, some experiments, such as specific absorption rate (SAR) and hyperthermia cannot be done on human beings due to the need to monitor variation of power intensity and temperature inside tissues. Employing live human participants for testing devices exposes the entire test procedure to several inherent uncertainties, such as respiratory movement, cardiovascular vibration and variable skin humidity in addition, of course, to the safety concern of the new devices. Also, application of the devices and systems on human subjects or human-related materials is a serious ethical issue where the researchers must receive an ethical clearance from proper authorities, and it can be difficult to reasonably estimate and investigate the level of risks from various scientific, physical and psychological aspects beforehand. Thus, the utilization of artificial tissue emulating (ATE) phantoms is much beneficiary for the testing purpose of a device or system.

Emulating a human organ or tissue with an exact substitute requires an accurate artificial phantom that is expected to have several features: anatomically realistic, dielectric precision across the band of interest and long lifetime. The anatomic distribution of various tissue layers in the phantom has to be accurate to analyse the real scenario. This is especially important if the application is heavily dependent on the constitution of the body part, like implantable devices. The electrical properties of the phantoms are essential to mimic the real situation. The properties of the ATE materials are compared with those of actual tissues over the band of interest. Lifetime of the phantom is particularly important for the repeatability of experiments and continuation of research. The shape and tissue distribution as well as the dielectric properties of the appropriate phantom has to be stable over a sufficiently long time.

The electrical properties (permittivity, $\varepsilon_r$ and conductivity, $\sigma$) of various tissues of the human body are vastly different according to their types. The properties also vary in wide range with respect to operating frequencies. Several researchers have systematically measured the properties for various tissues of the body [1]-[5]. Table I lists the range of values of the electrical properties (dielectric constant, or relative permittivity, and conductivity) of fifteen different main tissues of the human body for the frequency range of 500 MHz to 10 GHz as extracted from [3], [4]. It is noted that since various studies are not taken from the same samples or subject and as the dielectric properties of some tissues vary from individual to another and with age, the values presented in [1]–[5] are slightly different from each other with a reasonable tolerance level [6]. As a result, whilst various human ATE materials and body phantoms have been designed and reported by idealizing different sources, they in fact represent human tissues in different scenarios [7], [8]. Acknowledging this, an effort is made in this review to address different state-of-the-art ATE materials and phantom types for various operating frequencies, and fabrication procedures in order to have a better understanding of the pros and cons of various ATE phantoms which leads us to develop superior version of artificial human body substitute for various applications.

## II. Applications of ATE Materials

### *A. Dosimetric Studies*

The tremendous advancements in using handheld wireless microwave devices increased the public health concern from the emission of electromagnetic energy while placed in close proximity to human body, especially the head. For that reason, investigations have been done to analyse the SAR of different devices in various positions to estimate the amount of power absorbed by the human body. In order to estimate and verify the radiation safety of the devices, laboratory tests are mostly performed on ATE phantoms. Fig. 1 exhibits some reported experiments to evaluate the exposure of the human head to the near field of radiating antennas and devices close to the head [9], [10]. Most of the utilized ATE phantoms are anatomically non-realistic, but having average dielectric properties of the overall body part for the operating frequency band, e.g. SAM (Specific

Anthropomorphic Mannequin) phantom. However, in case of SAR estimation in the human body, the worst case evaluation might be more important than exact SAR values.

Typical methods of estimating SAR include approaches based on the measurements of electric field, magnetic field and temperature distributions inside ATE phantoms. SAR can be estimated by measuring the electric intensity using an electric sensor and applying commonly known correlating formulas [9, 10]. Due to the compression of the wavelength inside high permittivity phantoms, the operating frequency and resolution of SAR inside the phantom mostly depends on the compactness and performance of the sensor. The magnetic field based SAR test can be performed by using similar platforms as the electric field based methods [11]-[13]. Since measurements relying on the sensors require movements inside phantom, liquid and gel type ATE materials are mostly suitable for these SAR estimation methods. Owing to the fact that the temperature rise and SAR are proportional for a particular electromagnetic absorption, SAR can be calculated by measuring the distribution of temperature rise [14, 15], which can be measured by using thermal sensors or cameras. Thermal sensors can be non-perturbing temperature probes (using all types of ATE phantoms), should be planted inside the phantom prior to SAR estimation (for liquid, gel or semisolid ATE phantoms), or even before the phantom fabrication (solid ATE phantoms). Thermographic cameras, which can only measure surface temperature, require ATE phantoms that can be separated into different halves/quarters thus limiting the use of solid phantoms for SAR estimation inside them [16, 17]. However, the advanced and sophisticated thermal cameras are able measure the distribution of temperature rise even inside the phantoms, allowing the use of any type of phantoms for SAR measurements [18]. In temperature based SAR estimation method, the thermal properties of the ATE phantom are also considered to accurately measure the SAR values [19].

### *B. **Microwave Imaging***

ATE phantoms are extensively used in validating various microwave imaging algorithms and systems. Microwave imaging techniques have been widely investigated as a possible diagnostic tool in detecting the abnormalities (e.g. cancerous tumours, haemorrhage etc.) of various human body parts including breast, head, limbs and torso. Test beds of the imaging systems extensively require ATE phantoms with realistic tissue properties and distributions. To meet the increasing demand of realistic body mimicking parts, a lot of phantoms have been developed that emulate the actual environment. Fig. 2 depicts the application of breast and head phantoms in different microwave imaging experiments [20]-[24].

### *C. **Implantable Devices***

ATE phantoms play crucial part in the validation of wireless implantable devices that are utilized for collecting electro-physical data of human body/organs for monitoring and diagnosis, the treatment of various neurological diseases and controlling of prosthetic devices applications [25]-[27]. ATE phantoms allow the researchers to analyse the performance of the devices and overall system before clinical trials. The normal case scenario can be implemented on ATE phantoms in order to get an insight of the potential health hazards (e.g. heat generation, wireless power failure, device failure etc.) that might be incurred by the implantable circuitry. Consequently, implementable devices with biocompatibility, optimal operation, reliable data transfer and superior safety can be ensured leading to the advancements of such devices in medical and neurological sectors. Several reported applications of ATE phantoms for implantable devices are demonstrated in Fig. 3.

### *D.* ***Body-worn Sensors & Devices***

Recent developments of textile and wearable electronics have led the evolution of miniaturized body-worn devices for a wide range of applications in healthcare, defence, entertainment, sport and space, which can be divided into two main kinds: on-body and off-body communications. On-body communication mainly refers to the wireless interaction between two wearable devices or sensors, whereas off-body communication is defined as the communication between the wearable devices and base unit or mobile device placed in a short range area [31], [32]. Ideally, wearable devices are required to exhibit environment independent performance. However, the designs of the devices largely depend on the propagation channel conditions, such as the existence of strong multipath around the human body. ATE phantoms are widely applied to analyse the wave propagation of the body-worn devices and their overall performance in a repeatable manner [33, 35]. Fig. 4 shows the applications of ATE phantoms for verification of wearable devices and systems.

### *E.* ***Other Applications***

There are several other application where ATE phantoms are widely used for the verification purpose. The electrical and thermal studies of heat deposition by electromagnetic hyperthermia apparatus require highly sophisticated ATE phantoms. Most of the early ATE materials and phantoms are actually designed for hyperthermia applications [36]-[38]. In another application, the performance of designed antennas for handheld devices or mobile phone applications needs to be evaluated in the presence of ATE phantoms (Fig. 5(a)). Microwave radiometry for measuring the internal temperature profile of human tissue [40], [41] and non-invasive monitoring of blood glucose levels [42] also need to be verified by equivalent phantoms (Figs. 5 and 6).

## III. TYPES OF ATE MATERIALS

ATE materials can be classified in various types from different perspectives. Depending on the electrical properties, they can be of low-water-content and high-water-content ATE materials [1], [43]. While that classification gives a useful information on the suitability of a certain ATE type to emulate a specific tissue, it does not provide any opinion on the benefits of using a specific type of ATE material over other types. In order to analyse the pros and cons of different types, the ATE materials are divided into four kinds according to their physical appearance.

### *A. Liquid ATE Materials*

Tissues with high water content, such as muscle and main brain parts, exhibit high electrical properties. One way to closely mimic those tissues is to use water, which is usually utilized in ATE materials as the main source of high permittivity, as the basis of liquid mixtures. The main advantage of liquid ATE materials is the easy preparation by mixing different components according to an appropriate recipe. The preparation procedure of several liquid ATE materials is discussed in [44]. Fig. 7 illustrates the generalized flow chart for the fabrication of a muscle equivalent ATE material. A gentle heating is applied at the early fabrication stages to warm up the solution and enable creating the required solutions faster, although a keen observation to the temperature has to be maintained to reduce the evaporation of water caused by the heat.

To fabricate ATE materials of low permittivity, like fat and bone, a low percentage of water is used in the recipe. Several oil-in-water emulsions are reported in the literature for low-water-content ATE materials [44]. Another technique is to mix salted water with various nonionic surfactants (e.g. Triton X-100) [45].

The ATE liquid mixtures have some inherent disadvantages. It is difficult to maintain consistency of the material properties, as the materials tend to dehydrate, which drastically alters the relative permittivity and conductivity. Moreover, the mold growth might spoil the material by changing its electrical properties.

The moist materials are kept in shape using defining containers. Thus, a difficulty is imposed by the container at the time of emulating invasive or in vivo tests, such as implantable radio frequency components. The direct estimation of SAR on the phantom surface is also difficult to attain. In microwave imaging, for example, if the containers are not built from a material of suitable electrical properties, they are more likely to produce high scattering, which may jeopardize the validity of the experiments.

The components used to build liquid-based ATE materials are not usually dissolvable. Thus, to attain the desired electrical properties, the solution should be stirred properly until it becomes homogeneous. However, the used components in the solution tend to get separated forcing stirring the solution again; although early separation of the components exhibits minor changes in the electrical properties [46].

### *B. Gel ATE Materials*

Gel or semi-liquid ATE materials are more solid of course than liquid materials, but cannot hold its figure independently without a container or shell. These types of mixtures overcome the component separation problem of the liquids and sustain homogeneity over a long period. The fabrication procedure of these ATE materials are also straight-forward. A flow diagram of gel type muscle material as mentioned in [47] is depicted in Fig. 8. It is seen that the mixing is controlled using various rotation speeds as the mixture gradually becomes thicker. At the end of the procedure, the ATE material turns into a gel type substance, which is then poured into a phantom container.

Along with the advantage of the gel ATE materials, they demonstrate the rest of the weaknesses of liquid ATE materials.

Besides, there are some other issues. This type of ATE material take some time to form as gel's most undesirable feature is its variable setting time, which may extend to 1 day [47]. This variation may not be a serious problem to many applications, but it could disrupt the rapid phantom prototyping procedure and validation of the intended system.

As the ATE material is thick and prepared by continuous rotation, this type of material tends to trap air bubble during the fabrication. Hence, slow and consistent rotations should be applied at the end of the process to avoid the bubble formation.

### *C.* ***Semi-solid or Jelly ATE Materials***

Semi-solid or jelly ATE materials are capable to conform to any shape independently. This castable attribute of these materials is particularly beneficial to emulate the realistic situation of soft tissues, where the tissue layers have a particular form and pattern. Moreover, no osmosis is seen to be occurred between adjacent layers when multiple types of semi-solid ATE materials are placed one after another. As a result, the electrical properties are found to be stable in the investigations [48]. This feature allows the phantoms to be constructed in a multi-layered fashion, which reflects the anatomical structure of the human body reasonably well. The castable and non-diffusive natures are primarily because of the addition of right proportion of gelatin, agar or dough to ATE materials which also assist in adjusting the electrical properties of the final material. The fabrication process involves weighing the ingredients separately, heating of water by accordingly adding various components in different temperatures and finally mixing the preservative elements before cooling down to room temperature for storing or experimental purpose. A generalized graphical diagram of the fabrication procedure of semi-solid ATE materials is illustrated in Fig. 9 following the procedure explained in [49].

The semi-solid ATE materials are relatively low-cost and are able to mimic human tissues over a wide frequency band. However, they also have some limitations. Although this type of material is useful in medical imaging purposes, experiments which require invasive measurements (e.g. SAR measurements, implantable devices, hyperthermia etc.) become complicated with this tissue type as altering measurement position results in deformation of ATE layers.

Unlike the liquid and gel ATE materials, semi-solid materials are usually neither reusable nor dielectrically adjustable. For example, the electrical properties of liquid and gel materials can be altered by adding water or oil components. Consequently, the electrical properties can be adjusted. Following this procedure, the liquid and gel ATE materials can be used to build phantoms of different parts of the body (e.g. head, arm, leg etc.). This feature is usually not present in the semi-solid ATE materials. Nonetheless, the materials also tend to dehydrate and it is hard to preserve the materials over a long period of time [45], [49].

### *D.* ***Solid ATE Materials***

Since solid ATE materials are not water-based, they overcome the hydration drawbacks of the liquid, gel and semi-solid ATE materials. For this reason, they are called dry ATE materials [14]. Those materials are built primarily from ceramic powders, which are available in wide varieties with a wide permittivity range [14]. Nonetheless, the ceramic materials are of low loss, which represents a challenge to building lossy materials to emulate the actual conductivity of many human tissues. However, the composition of ceramic powders with various conductivity enhancing materials provide the flexibility of separately controlling the permittivity and conductivity values of the solid ATE materials [50]. An outline of the solid phantom fabrication as mentioned in [19] is shown in Fig. 10. To fabricate solid materials special production equipment is required due to the need for high temperature for material blending and high pressure is required for injection molding purpose. The specialized ceramic materials and instrumentations make the whole fabrication procedure more expensive than the rest of the ATE materials. Moreover, specialized adhesive materials are used with the ceramic powder for removing the air gaps between adjacent ceramic pieces. Those adhesive inclusions are difficult to use and make the ATE phantoms difficult to cut or reshape. In order to overcome this problem, soft and dry ATE materials are introduced utilizing silicone rubber with carbon fiber [51] or carbon nanotubes [52], which can be used to develop lightweight solid ATE materials whose specific density is less than 1. Whilst experiments that require invasive measurements can be problematic with the solid ATE materials, this type of ATE material can sustain over a long period of time without changing its shape. Moreover, a temperature dependency of the dielectric properties is observed, which may be beneficiary for some applications, like hyperthermia.

## IV. ATE PHANTOMS

In this section, the ATE phantoms reported in the literature are described according to the human body parts for both narrow and wide frequency range. Although a lot of anthropomorphic [53] and realistic [54] phantoms of human body parts are built for CT scan application, their electrical properties are not taken into account, and thus CT scan phantoms are not included in this section.

### *A. **Head Phantoms***

The head is one of the most complicated structures of the human body. Mimicking the human head is a challenging task as the head includes a lot of different tissues with complex distribution. To simplify the situation and ease the fabrication process, many researchers have utilized homogeneous ATE material, which equivalently represents the overall dielectric properties of the real head. In the narrow frequency arena, homogeneous liquid head phantoms are mostly prepared from sugar-water-salt based solutions with various preservatives [3], [10], [55], [56] or using readymade commercial liquids as ATE materials [9], [39], [57], [58]. Along with sugar-water-salt based solutions [10], a mixture of propylene glycol and deionized water has shown broadband performance in homogeneous liquid phantoms [59]. The dielectric properties of brain equivalent materials in those phantoms are calculated by averaging the properties of gray and white matter tissues [10], [56], [59]. The container of the liquid is made from low permittivity polyster or Polyvinyl chloride (PVC) materials. It is found that lumping of polymerizing materials during the fabrication procedure of liquid ATE materials can be resolved using water-soluble hydroxyetheylcellulose (HEC). This viscosity increasing agent also enhances the bacteria resistance of the solution and increases the lifetime of the phantom [10].

Gel head phantoms are basically agar based in both narrowband [30], [60] and wideband [61]; although polymer based substances are also utilized to attain intended properties for narrowband application [37]. Semi-solid head phantoms are more common in broadband applications. These phantoms are built from several homogeneous ATE materials which are then utilized to fabricate heterogeneous/layered semi-solid phantoms [8], [22], [23], [49], [17], [62]-[66]. The phantoms are mostly agar based to maintain the phantom's shape. Several gelling agents, like TX-151 is used in several phantoms to increase the viscosity of the materials [8], [41], [17], [62], [65], [66].

There are very few solid head phantoms reported in the literature. Fig. 11(a) depicts a photograph of the solid head phantom reported in [14]. The comparatively more sophisticated fabrication procedure and expensive materials of solid phantoms than the other ATE counterparts might be the underlying reason. The solid head phantoms are mainly designed for narrowband applications [2], [14], [50], [67], [68]. However, a broadband solid ATE material is reported for broadband application in the ultra-high frequency (UHF) band [69]. In terms of accuracy of the tissue distribution, most of the phantoms are anthropomorphic. Several multilayer stylized human head phantoms with realistic ATE materials are also found in the literature (Fig. 11(b)). Recently, several studies reported three-dimensional (3D) printed head phantoms with semi-solid homogeneous and multi-layer realistic tissue compositions [23], [49]. Fig. 11(c) illustrates the three-dimensional 3D printing selective laser sintering (SLS) process. However, filling up the intracranial portion with semi-solid materials and 3D fabricated molds is a challenging task especially as the human head has a complex structure. Details of the fabrication process and layer are demonstrated by using a photographic sequence in Fig. 12.

It is noted that most of the reported ATE materials are proposed for specific applications. Analyzing them reveals the suitability of various phantoms/materials in verification of different systems or devices. The study of SAR measurements and hyperthermia technique verification are largely performed using homogeneous liquid and gel head phantoms. Solid head phantoms are also used for SAR measurements, especially where there is a flexibility of thermographic measurements [2], [14], [50], [69]. Otherwise, wearable communication systems mostly use solid head phantoms. Homogeneous gel and heterogeneous semi-solid ATE phantoms are popular in implantable device experiments. On the other hand, microwave imaging systems are validated using homogeneous gel and heterogeneous semi-solid head phantoms.

It is found that exterior of most of the liquid phantoms are constructed from low-loss plastic or polyester type materials. Although the shelf lifetime of the phantoms is one of the important issues in the fabrication strategy, descriptions are found to be insufficient for most of the reported works. An easily developable liquid tissue has been reported to attain over one year long lifetime [44]. Some gel type ATE materials have shown dielectric consistency over 40 days [61]. However, most of the semi-solid ATE materials are highly vulnerable to dehydration due to their lower water content, they limit the safe testing time and have comparatively short lifetime. Some semi-solid ATE materials have demonstrated reasonable changes (2-5%) over 4 weeks [22], [63], [64], [70], [71]. The fabricated head phantoms are summarized in Table II.

### *B. **Breast Phantoms***

All the breast phantoms reported in the literature are proposed for the validation of microwave based imaging systems aiming at breast cancer detection. Oil and gelatin emulsion based semi-solid phantoms are very popular due to their rapid fabrication procedure and possible ease in multi-layer formation.

Several oil-in-gelatin based anatomically realistic breast phantoms are also reported along with few hemispherical breast models. These breast phantoms are heterogeneously constructed according to the MRI images of actual human breasts by using 3D printed molds [7], [48], [75], [76]. Fig. 13(a) presents the different parts of MRI-derived breast molds and Fig. 13(b) depicts a hemispherical breast model indicating different types of fabricated ATE materials. Nevertheless, the verified CT images presented in [48], exhibits the fabrication accuracy of the breast phantom. However, they have limited lifetime, which lasts up to 8 weeks [77]. For this reason, recently, Triton X-100 and water based liquid phantoms have become increasingly popular [45], [78]-[80]. This type of mixture is relatively stable for around 1 year [45]. Moreover, the dielectric properties of the mixtures can be varied from low to high-water-content tissue properties. Thus, this type of mixture is used as the filler of 3D fabricated breast phantoms made from frequency dispersive solid materials (shown in Fig. 13(c)) [80].

The reported mimicking breast tissues and phantoms are mostly designed for wideband operation [7], [20], [45], [48], [77], [76], [79]-[89]. Solid breast phantom are also noted [90], [91], however, they are not widely adopted due to their non-dispersive dielectric properties. An example of solid dielectric breast phantom is illustrated in Fig. 13(d). Various breast phantoms are listed in Table III in accordance to their reported frequency bands.

### *C. **Limb Phantoms***

The limb phantoms are designed for various broadband applications, such as electrical impedance tomography, hyperthermia and wearable communication systems [1], [38], [44], [46], [93]-[96]. The range of frequencies of those phantoms extends from a few kHz to tens of GHz depending on the application. According to their types, most of the limb phantoms are constructed homogeneously using tissue/skin equivalent materials [46], [95], [96]. However, some reported limb phantoms have heterogeneous structures [1], [38], [93]. The lifetime of the limb phantoms are reported to be in 2 to 4 weeks [95], [96]. Examples of several limb phantoms are illustrated in Figs. 14(a-c). In Table IV, the reported limb phantoms of the literature are listed according to their designed frequencies.

### *D. **Torso Phantoms***

Torso phantoms are mostly constructed for hyperthermia and electromagnetic radiation (SAR) studies of various organs [20],

[97] and over the whole body [37], [47], [66], [99]-[102]. However, some phantoms are reported for ultra-wideband applications [102], [103]. Several heterogeneous torso phantoms are reported for the validation of various monitoring techniques [42], [98]. The recipes of skin, fat, muscle and bone reported in aforementioned ATE phantoms can be easily adopted to construct a torso phantom. Agar based semi-solid ATE materials are seen to maintain soft body tissue equivalent mechanical strength [104]. Photographs of various ATE human torso are demonstrated in Figs. 15(a-c), whereas a summary of them is listed in Table V.

## V. FEATURES AND FUNCTIONS OF INGREDIENTS IN ATE MATERIALS

The ATE materials described in the previous sections can be applied to build multiple types of phantoms. For example, the muscle can be found in all over the body, thus the muscle tissue mimicking materials described for head is also applicable in fabricating phantoms for other body parts. The ingredient composition of an ATE material defines its physical characteristics and dielectric properties. Altering the composition of the utilized ingredients wisely can be used to control the dielectric properties over a wide range which can be very useful in realizing other types of human tissues [45]. Moreover, the physical and mechanical characteristics can be altered by introducing some new ingredients or decreasing some existing ones. For these reasons, a proper understanding of the reasons behind applying different materials is very important for researchers working on phantom design. From the numerous materials used for the fabrication purpose of the various ATE materials and phantoms discussed above, the features and functions of the vital materials that define the material characteristics are listed in Table VI.

## VI. FUTURE TRENDS OF ATE PHANTOMS

With the ever increasing demand for new devices and systems that interact with the human body for different applications, it can be expected that the current efforts for the development of ATE phantoms will continue to evolve. Looking forward, there are several issues that need to be addressed or implemented by the researchers for the accurate validation of body-centric *in vivo* and *in vitro* wireless systems and devices.

### *A. **Age-dependent Phantoms***

With age, the tissue properties gradually change because of the variation in the water content and organic composition of human tissues. Although the dielectric properties of tissues are widely studied, sufficient data are not available for age-dependent human tissues analysis. Hence, several researchers have systematically analyzed the age-dependent dielectric changes of rats and pigs, and proposed numerical solutions to estimate the age-dependent inherent tissue properties [113]-[117].

Several researchers have built numerical models of age-dependent human body models by morphing deformation of an MRI-evolved adult human model [57], [118], [119]. Considering the age-dependent tissue properties and MRI-scan based numerical child models in dosimetric simulations reveals that on average children suffer from a higher radiation exposure of their brain parts compared to adults, owing to the anatomical differences between the models [121], [122]. This provokes some major concerns regarding the child safety [123]. Since the dosimetric results are context dependent, in order to consider a device to be safe for all ages, experimental investigations should be performed on ATE phantoms of different ages. However, age-dependent ATE phantoms are rarely reported in the literature [41], [60].

It is also found that the tissue distributions inside head of children are different from adults [114], [122], [124]. Scaling down the adult head models following [57], [118]-[120] can lead to a large number of uncertainties [125]. As a result, proper validation of wireless devices on children might be incorrect, which might be crucial in diagnostic, therapeutic or monitoring of patients in clinical applications. Hence, applications which involve children should be exclusively verified by age-dependent child ATE phantoms.

The aforementioned statements are equally true for the fetus models. A few MRI-scan based voxel models are numerically analyzed by various researchers using FDTD codes and simulation tools [126]-[128]. Although several electrical and thermal properties of the fetus can be found in the literature [129], [130], some of them are with uncertainty and in the absence of various actual fetal and particular pregnancy-related (e.g. placenta) tissue data, several assumptions are taken into account to predict their properties [126], [128]. However, pregnancy-specific ATE phantoms are yet to be introduced in dosimetry research.

### *B. **Thermal Properties***

In various applications, the main objective of the devices involves temperature performance rather than power deposition. Examples of those applications include electromagnetic diathermy, internal power density analysis, ablation, etc. Those systems demand ATE materials to provide internal temperature distribution equivalence of actual issues along with the equivalence of electrical properties [36]. Thus, the ATE materials of various tissues are additionally required to have thermal capacity or specific heat, and thermal conductivity with various thermal time constants or thermal diffusivity depending on the application. In order to measure or display the temperature distribution of a phantom in adiabatic conditions, ATE phantoms with reasonably large time constant are required so that adequate amount of time is available to perform measurements before the temperature distribution deteriorates from the actual values [37]. Several researchers have measured the thermal properties of the different tissue types, and suggested mathematical models to postulate the thermal properties of human tissues in various frequencies and temperatures [132], [135], [136]. It is noted that the specific heat and thermal conductivity values of the high-water-content tissues (e.g. muscle, cancerous tissues etc.) are higher than those of low-water-content tissues (e.g. fat, bone etc.). However, due to the increased fabrication and validation complexities, there are very few phantoms reported to address these issues in details [19], [135], [136].

In temperature related applications, such as breast hyperthermia, blood flow might play a significant role in the cooling mechanism. However, breast phantoms emulating the circulation of blood flow in thermoregulation and cooling process of the human body during hyperthermia or other therapeutic procedure are yet to be reported.

### C. *Thermal Dependencies of Dielectric Properties*

The electrical properties of human tissues vary in different temperatures mainly due to different water concentrations of tissues in different temperatures [45], [131]. Fabricating phantoms by relying on the electrical properties at a single temperature (usually room temperature) is not appropriate for the aforementioned temperature related applications. The interaction between the tissues and propagated electromagnetic waves, and the thermal reaction of those tissues to electromagnetic power accumulation cannot be realistically estimated. This is especially true in case of few therapeutic technologies like ablation and electromagnetic hyperthermia which works at temperature much higher than the room temperature. As a result, a significant alteration in focusing the targeted region, such as tumors in hyperthermia, might occur turning the entire system vulnerable to errors [38]. Although electrical properties of various human tissues are measured and studied with various numerical models, very few ATE materials with temperature dependent dielectric properties are found in the literature [38], [19], [133], [134].

### D. *Three-Dimensional Printing and CT Verification*

The ever-growing developments of 3D printing technology have enabled small-scale fabrication of complex structures. Although this technology has a lot of advantages in fabricating heterogeneous tissue distribution of ATE phantoms, the choice of fabricating substrates are very limited. This is why researchers have currently no options but to use exterior shells and molds to emulate the realistic heterogeneity [7], [49], [76]. This kind of fabrication is time-consuming. Moreover, it sometimes makes it hard to maintain tissue consistency due to unintentional fabrication errors and generated air bubbles. Hence, to verify the integrity and accuracy of the built phantoms, CT scans are required (Fig. 16) [48]. This verification is especially important for microwave imaging applications, as these scans also provide the possibility of comparing the ATE phantom with scan-derived numerical simulations. It is expected that in the near future, a range of substrates and/or new 3D printing apparatus will be available to help in the precise fabrication using semi-solid ATE materials according to a computer generated 3D model [137].

## VII. CONCLUSION

Artificial phantoms to emulate different human body parts are crucial in the performance and safety verification of wireless body-centric devices and systems. A review of the developed phantoms has been reported in this paper. Firstly, various applications of the artificial human phantoms are discussed to briefly illustrate the current use and importance of the phantoms. The tissue emulating materials are classified in liquid, gel, semi-solid and solid types. A qualitative analysis between various material types is performed to demonstrate the advantages of one kind over another. Moreover, the reported phantoms for various body parts, namely head, breast, limb and torso phantoms are discussed according to their characteristics and intended applications, whereas a summary of the phantoms are listed in tables according to their band of operation. The features and functions of the main ingredients utilized for the phantom fabrications are listed. The included information aims at making the fabrication process meaningful and thus helpful to design new and more efficient phantoms for current or future applications, which rely on the interaction between electromagnetic fields and specific human body parts. Lastly, limitations of existing phantoms are discussed revealing the future directions and required breakthroughs by coordinated and cooperative efforts from scientists of various disciplines for fabricating the state-of-the-art human phantoms ensuring rigorous safety and performance validation of microwave devices and systems.

Figure Captions:

Fig. 1 (a) Application of Specific Anthropomorphic Mannequin (SAM) phantom in dosimetric assessment system using a robot arm [9] (© [2009] IEEE), and (b) SAR measurements of mobile phone on a head-torso phantom [10] (© [2009] IEEE).

Fig. 2 Application of various phantoms on microwave imaging platforms: (a) a breast phantom [20] (© [2009] IEEE), (b) a heterogeneous hemispherical breast phantom [21] (reproduced courtesy of The Electromagnetics Academy), (c) a heterogeneous head phantom [22] (© [2014] IEEE), and (d) a 3D head phantom [23] (© [2014] IEEE).

Fig. 3 (a) Performance evaluation of an implantable antenna in a fabricated three-layered phantom [28] (© [2010] IEEE), (b) experimental validation of a bio-antenna inside a phantom [29] (© [2009] IEEE), (c) receiver implanted inside a head phantom, and (d) *In vitro* experiment of the implantable medical device [30] (© [2009] IEEE).

Fig. 4 The experimental setup for (a) radiation pattern measurements of an antenna with head and torso solid anthropomorphic phantom using optical fibre for off-body communication [34], (b) channel characterization of tissue phantoms for on-body communication systems [33] (© [2008] IEEE), and (c) on-body performance measurement setup of a triband antenna on a human phantom [35] (© [2012] IEEE).

Fig. 5 (a) Tests of a mobile phone antenna using a head phantom [39] (© [2012] IEEE), (b) placement of the anatomical head phantom in a microwave radiometry system [40] (© [2012] IEEE), and (c) Microwave radiometer with ATE phantom [41] (© IOP Publishing. Reproduced by permission of IOP Publishing. All rights reserved).

Fig. 6 Application of a four-layered tissue phantom (a) on top of a Patch resonator (b) in measurement setup [42] (© [2014] IEEE).

Fig. 7 An outline of liquid muscle ATE material fabrication.

Fig. 8 An outline of gel type muscle ATE material fabrication.

Fig. 9 A generalized diagram of gel type brain ATE material fabrication.

Fig. 10 A generalized flow chart of solid ATE materials for head phantom fabrication.

Fig. 11 (a) Solid human-head phantom reported in [14] (© [1993] IEEE), (b) multilayer stylized head phantom [22] (© [2014] IEEE), and (c) 3D SLS printing of head phantom reported in [23], [49].

Fig. 12 Different steps of the phantom fabrication (clockwise): (a) 3D printing process, (b, c) various 3D fabricated parts; head phantom after filling up (d) Dura layer, (e) CSF, (f) grey matter, (g) white matter, (h) cerebellum; (i) the halves of the fabricated phantom, (j) the whole head phantom [49] (© [2014] IEEE).

Fig. 13 (a) MRI-derived breast molds, showing the inner and outer skin molds and the interior glandular mold [7], (b) a built hemispherical breast model inserting tumor and glands [75] Copyright © 2011 Wiley Periodicals, Inc., (c) 3-D-printed breast phantom prior to filling the fibroglandular void regions with liquid [80] (© [2012] IEEE), and (d) solid dielectric breast phantom [91].

Fig. 14 (a) Life-size solid human head and hand phantom models [19] (© [1997] IEEE), (b) multi-layered semi-solid cylindrical limb phantom models [93] (© [1971] IEEE), and (c) semi-solid skin-equivalent phantom representing an arm and a hand [95] (© [2012] IEEE).

Fig. 15 Realistic human torso phantoms: (a) multi-postural ATE torso phantom at 0.9 GHz [105] (© [2001] IEEE), (b) commercial torso phantom [106] (© [2014] IEEE), and (c) rubber based muscle equivalent full sized realistic human phantom model [52] (© [2007] IEEE).

Fig. 16 CT images of (a) sagittal and (b) coronal cross-sections of heterogeneously dense breast phantom [48] Copyright © 2011 Wiley Periodicals, Inc. Features present in each phantom include a skin layer and areola (medium grey), fat (black), heterogeneous mix (medium grey), and fibroglandular (light grey) tissue.

Fig. 1

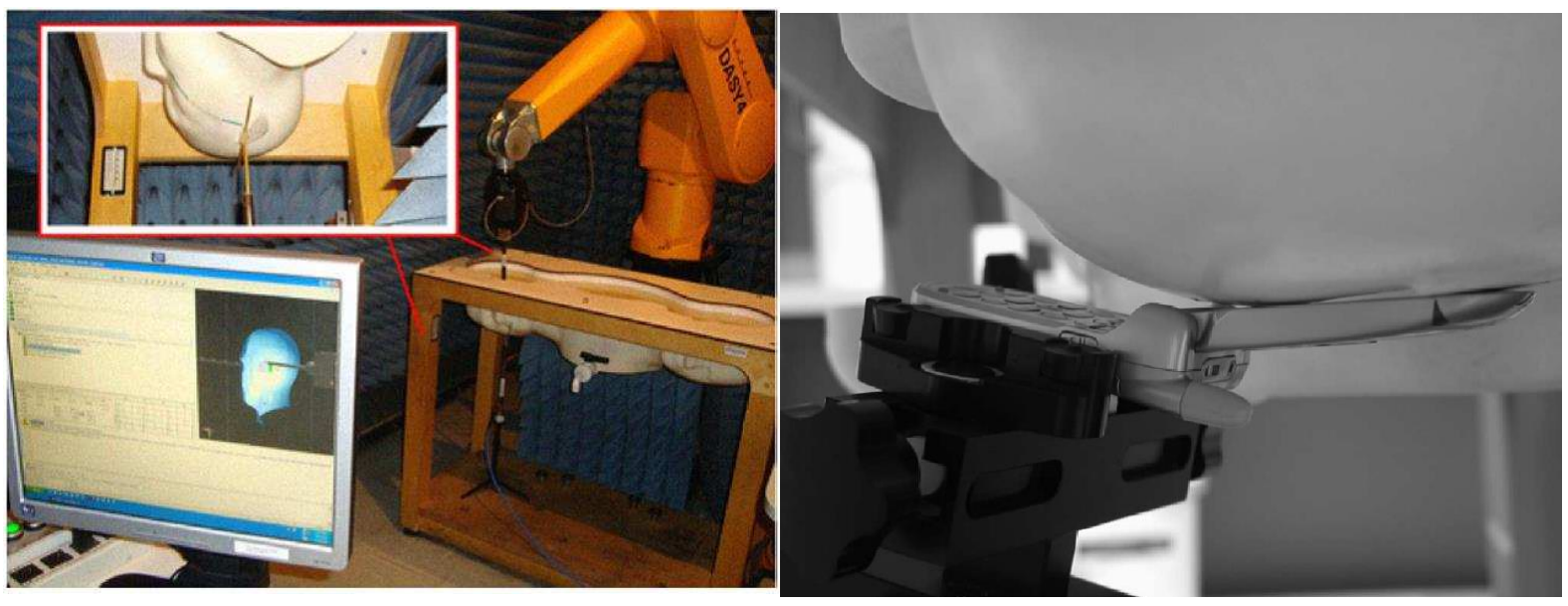

(a) (b)

Fig. 2

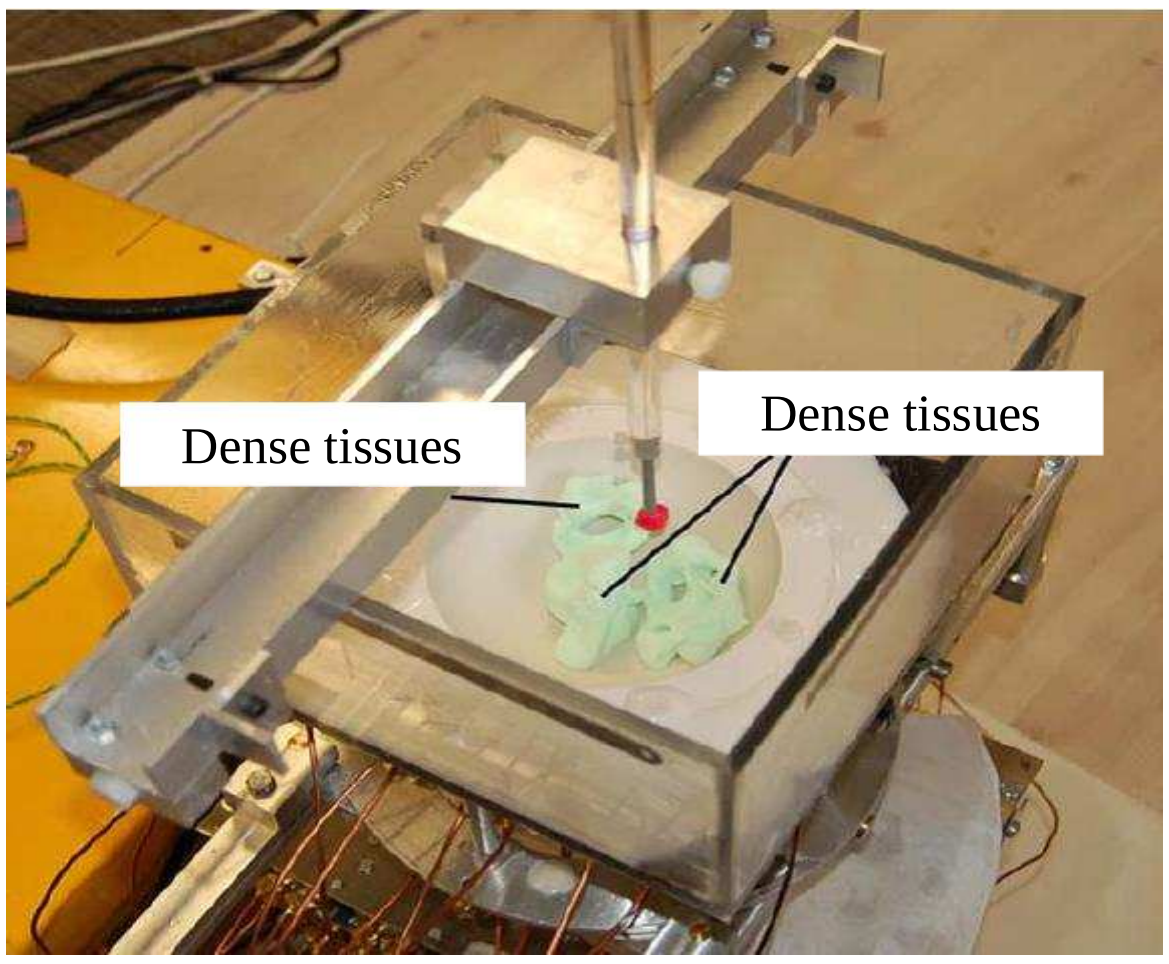

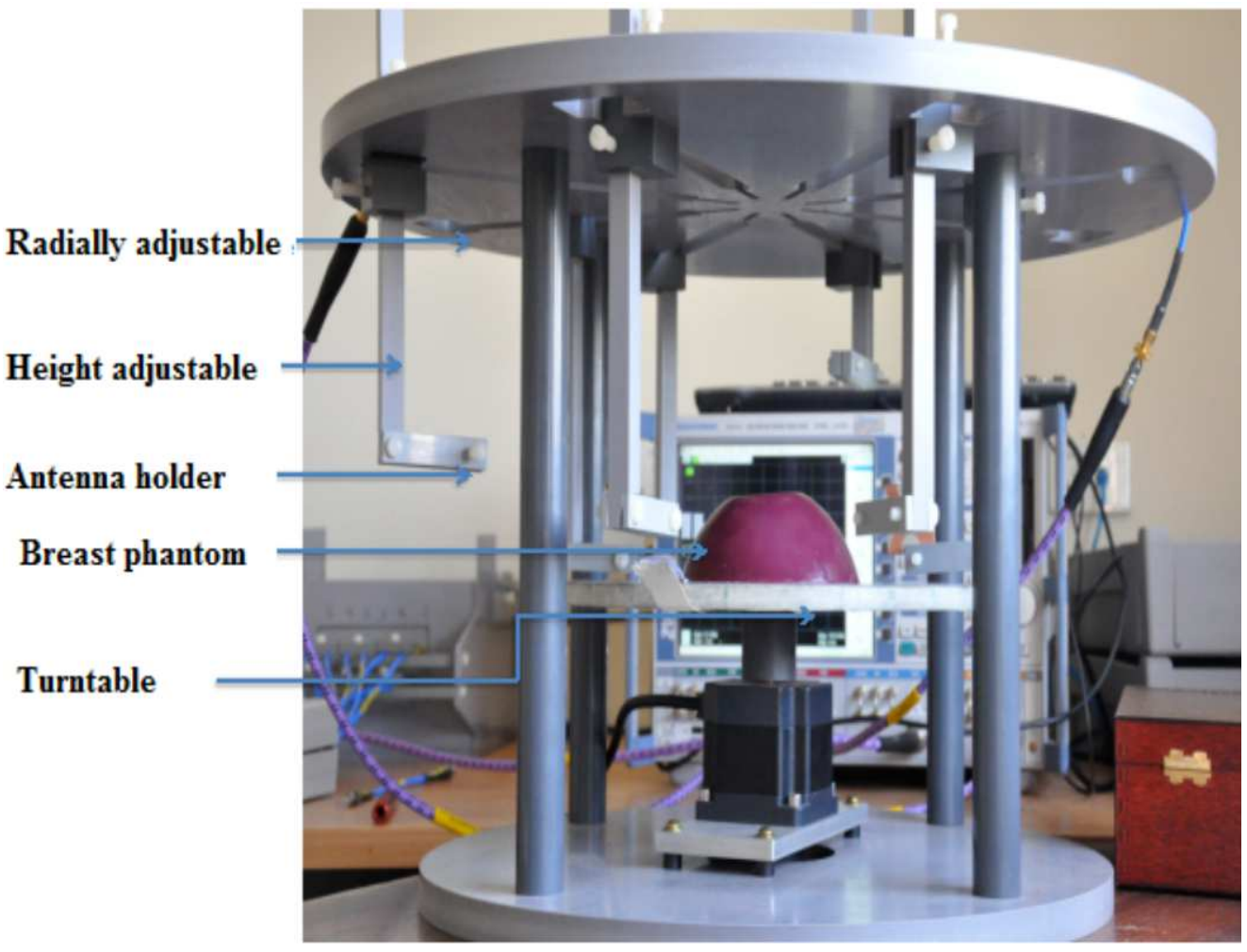

(a) (b)

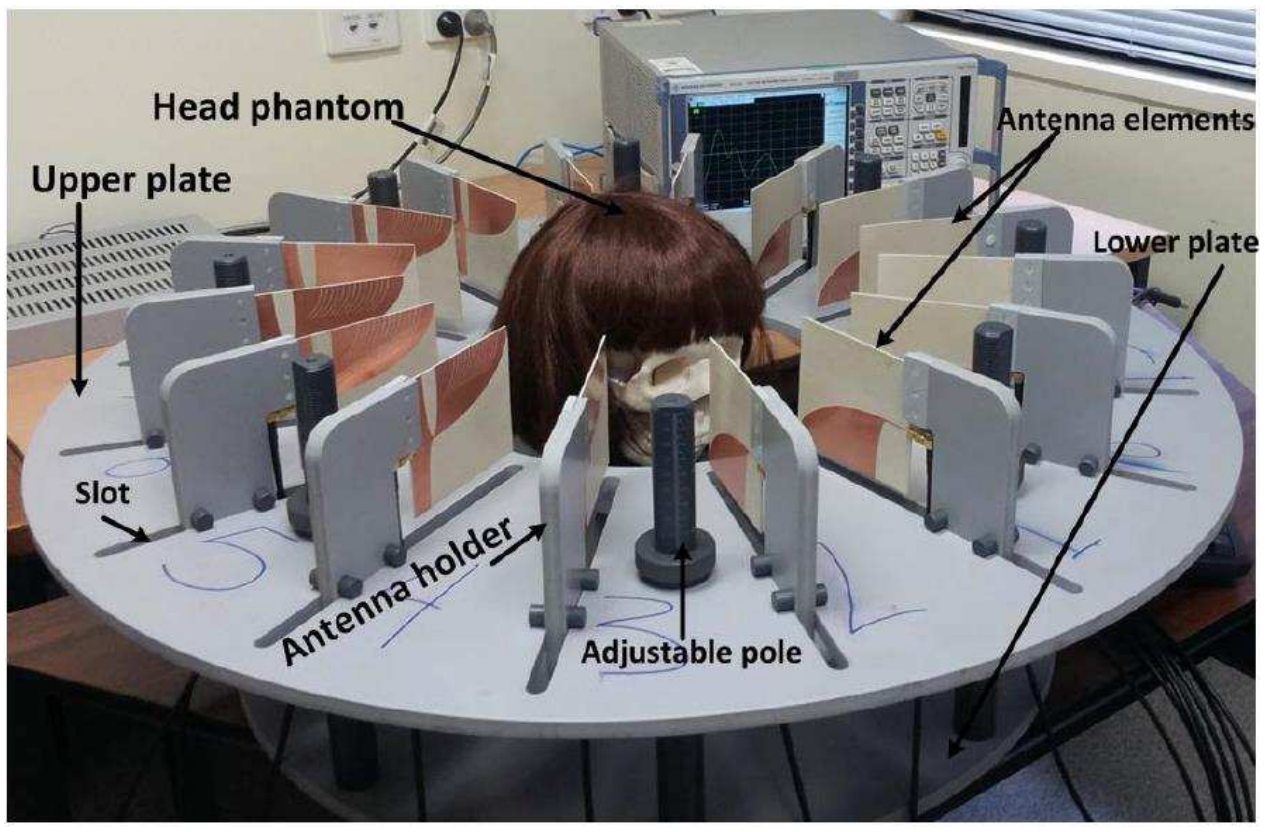

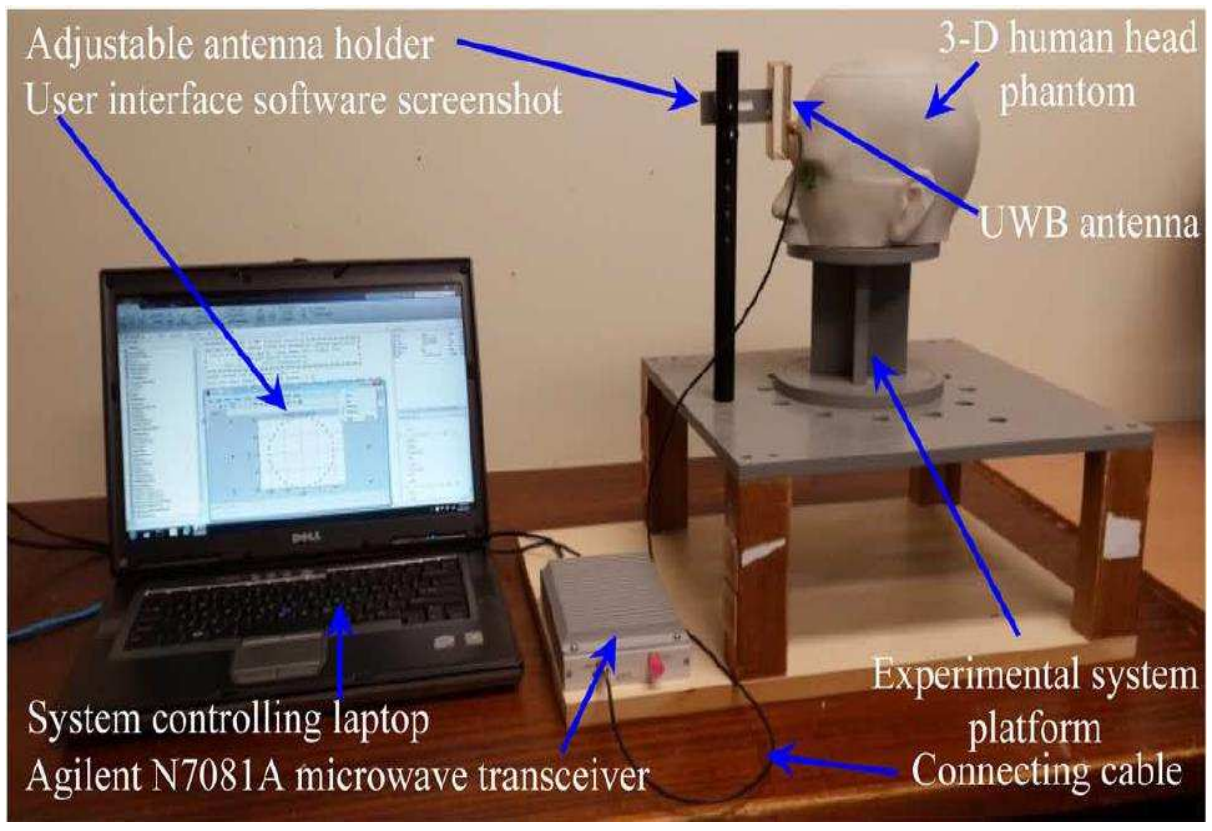

(c) (d)

Fig. 3

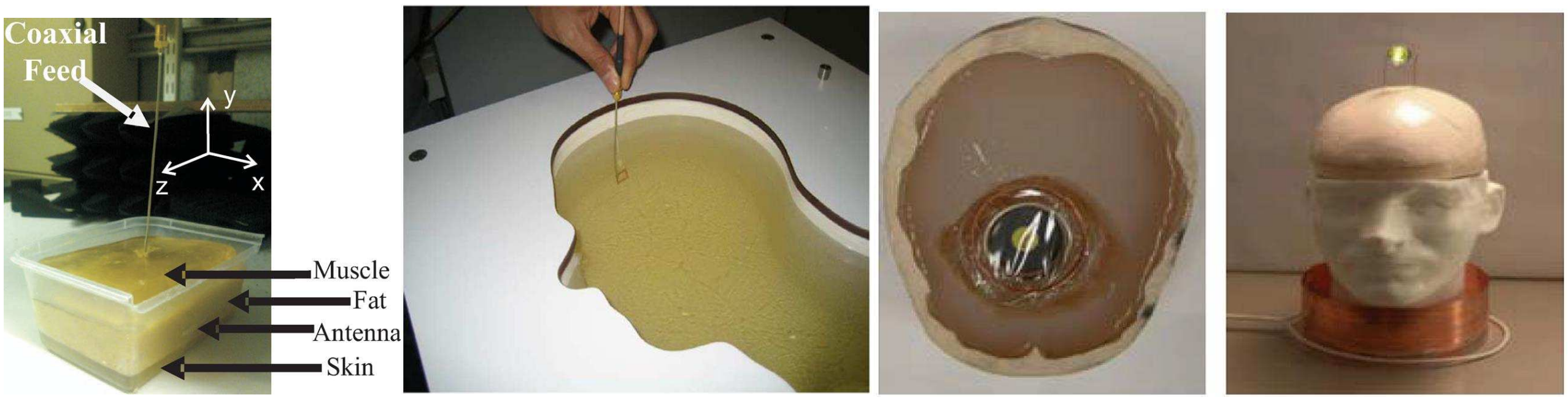

(a) (b) (c) (d)

Fig. 4

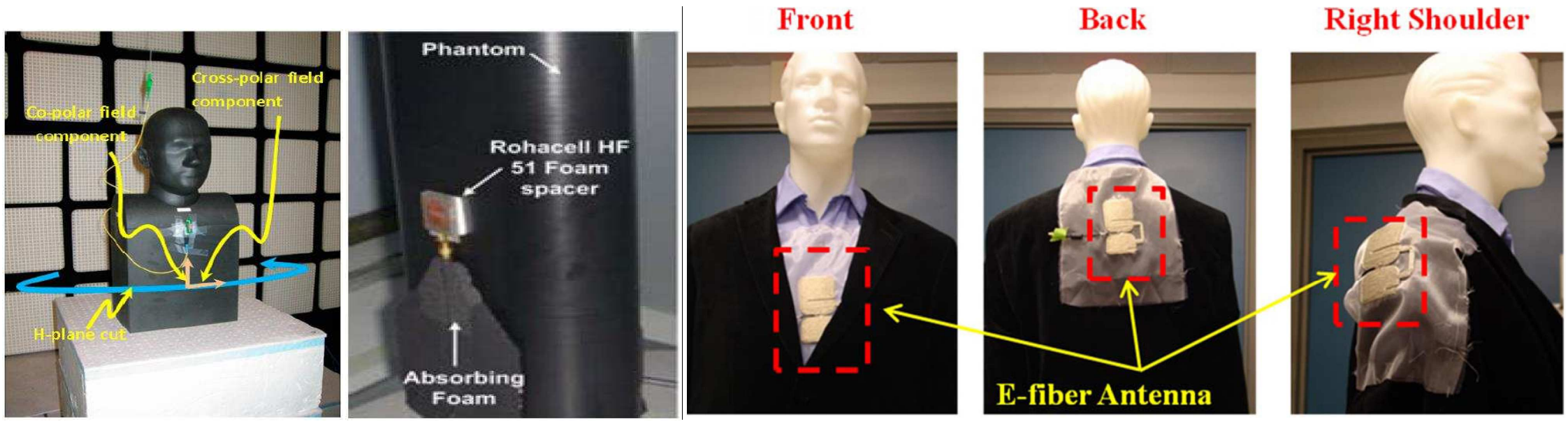

(a) (b) (c)

Fig. 5

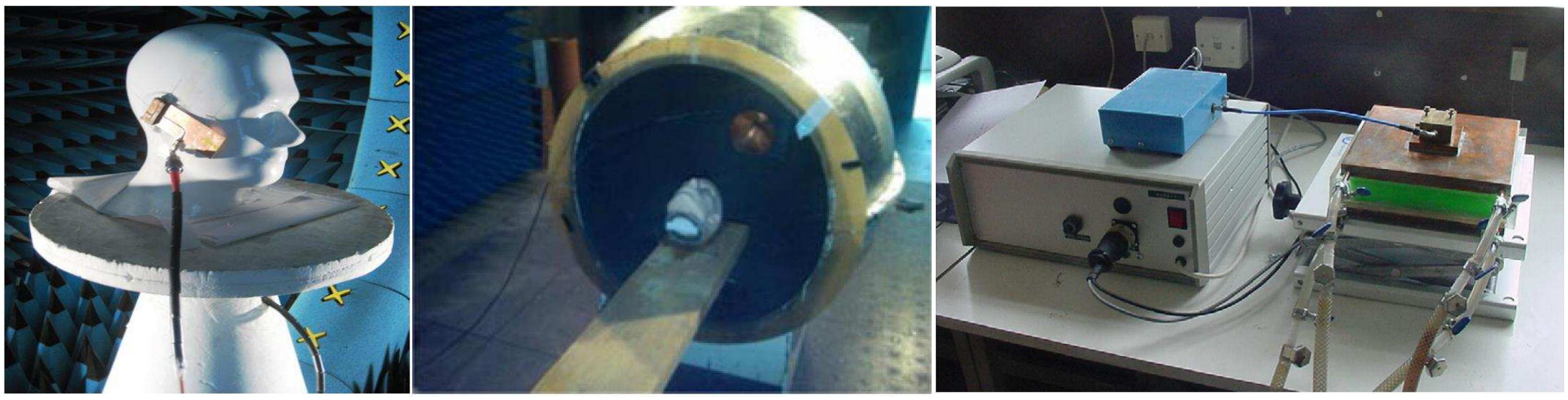

(a) (b) (c)

Fig. 6

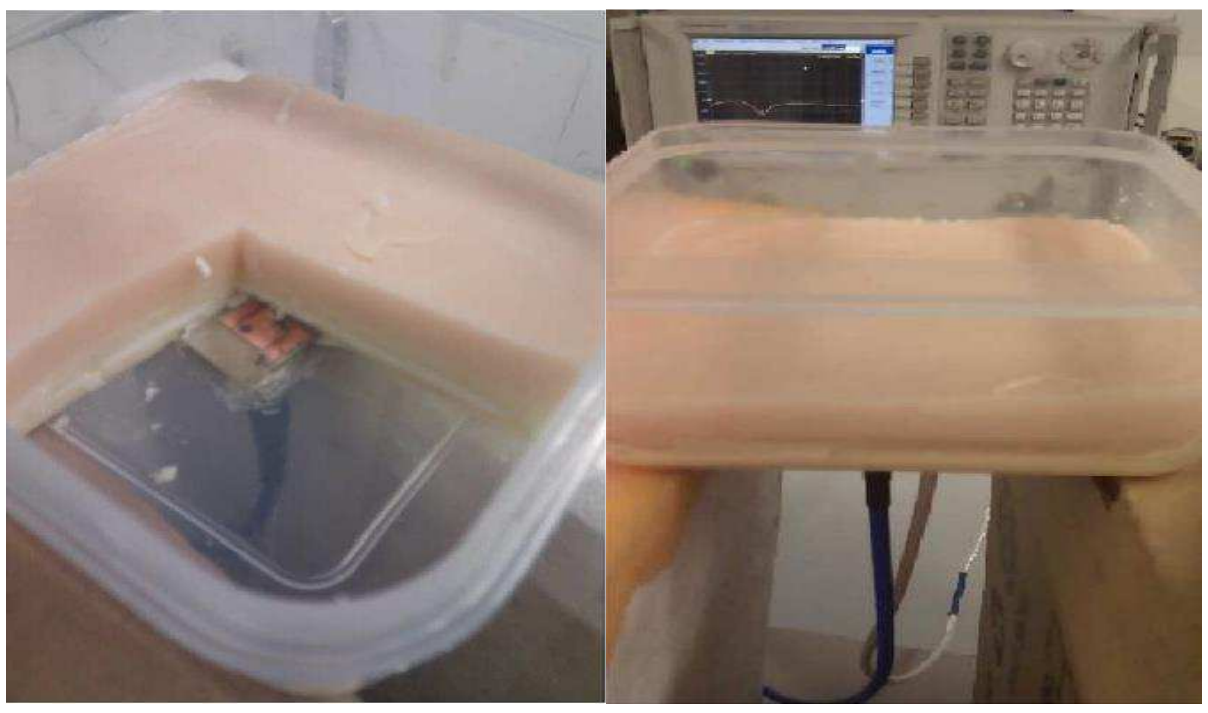

(a) (b)

Fig. 7

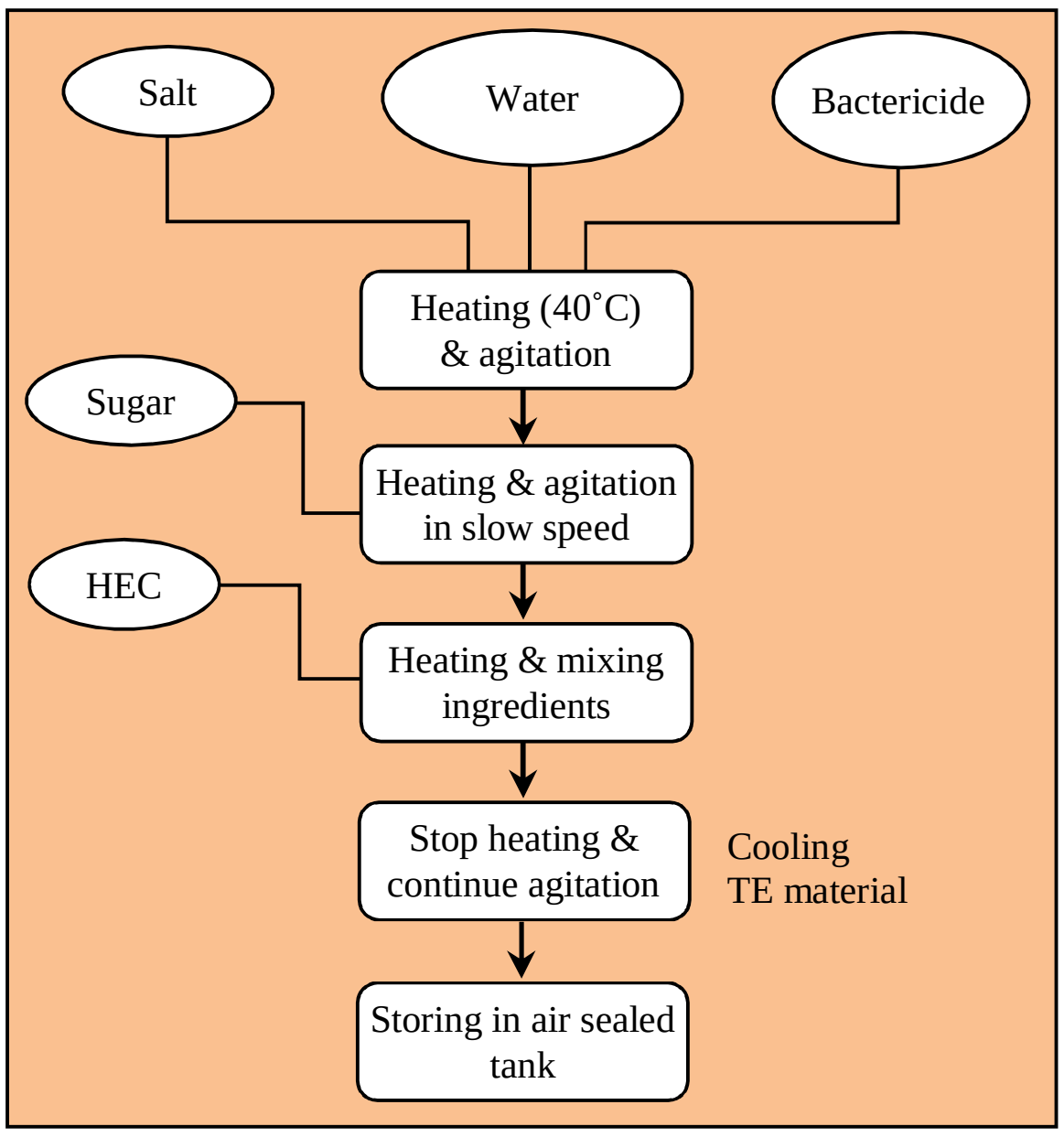
Salt
Water
Bactericide
Heating (40˚C) & agitation
Sugar
Heating & agitation in slow speed
HEC
Heating & mixing ingredients
Stop heating & continue agitation
Cooling TE material
Storing in air sealed tank

Fig. 8

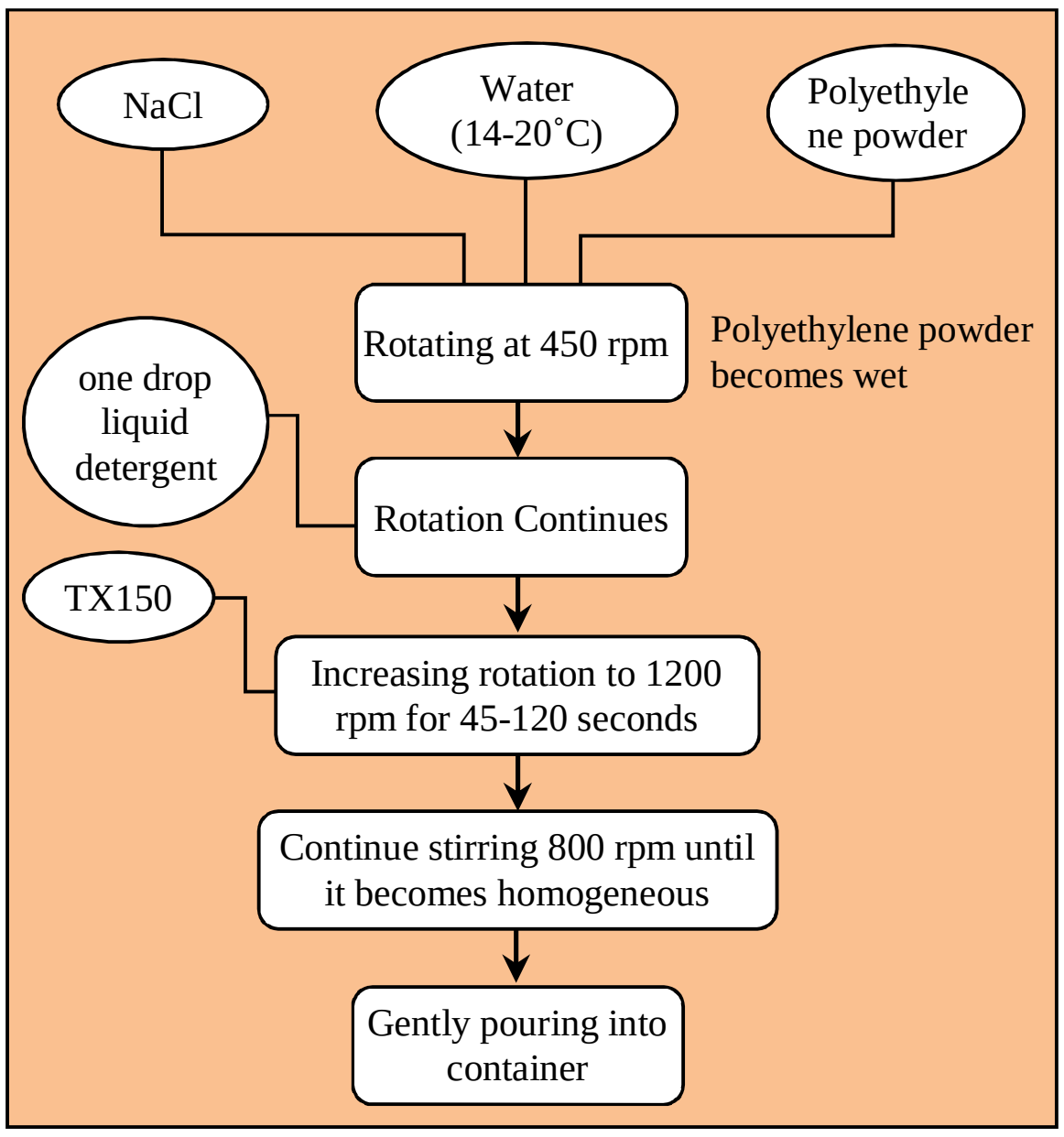
NaCl
Water (14-20˚C)
Polyethyle ne powder
Rotating at 450 rpm
Polyethylene powder becomes wet
one drop liquid detergent
Rotation Continues
TX150
Increasing rotation to 1200 rpm for 45-120 seconds
Continue stirring 800 rpm until it becomes homogeneous
Gently pouring into container

Fig. 9

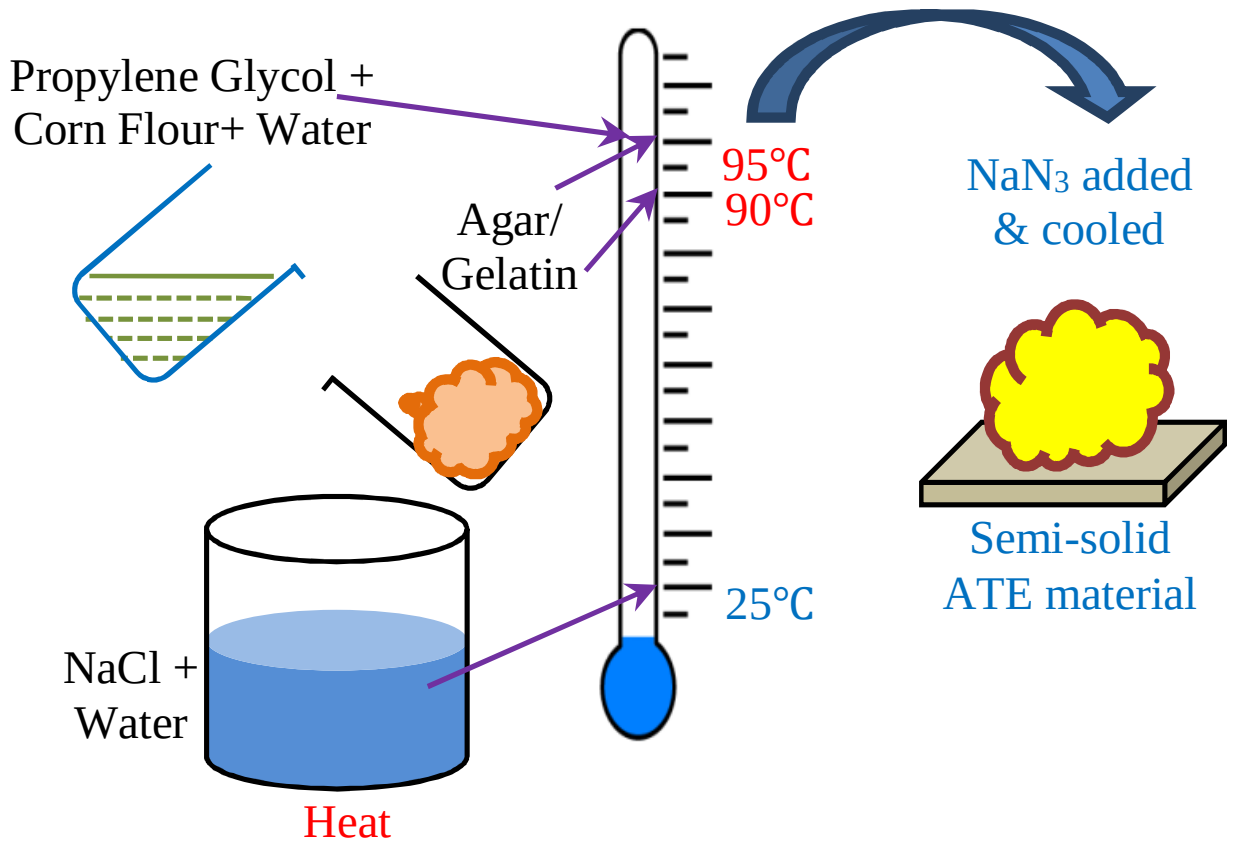
Propylene Glycol + Corn Flour+ Water
Agar/ Gelatin
95℃
90℃
NaN3 added & cooled
Semi-solid ATE material
25℃
NaCl + Water
Heat

Fig. 10

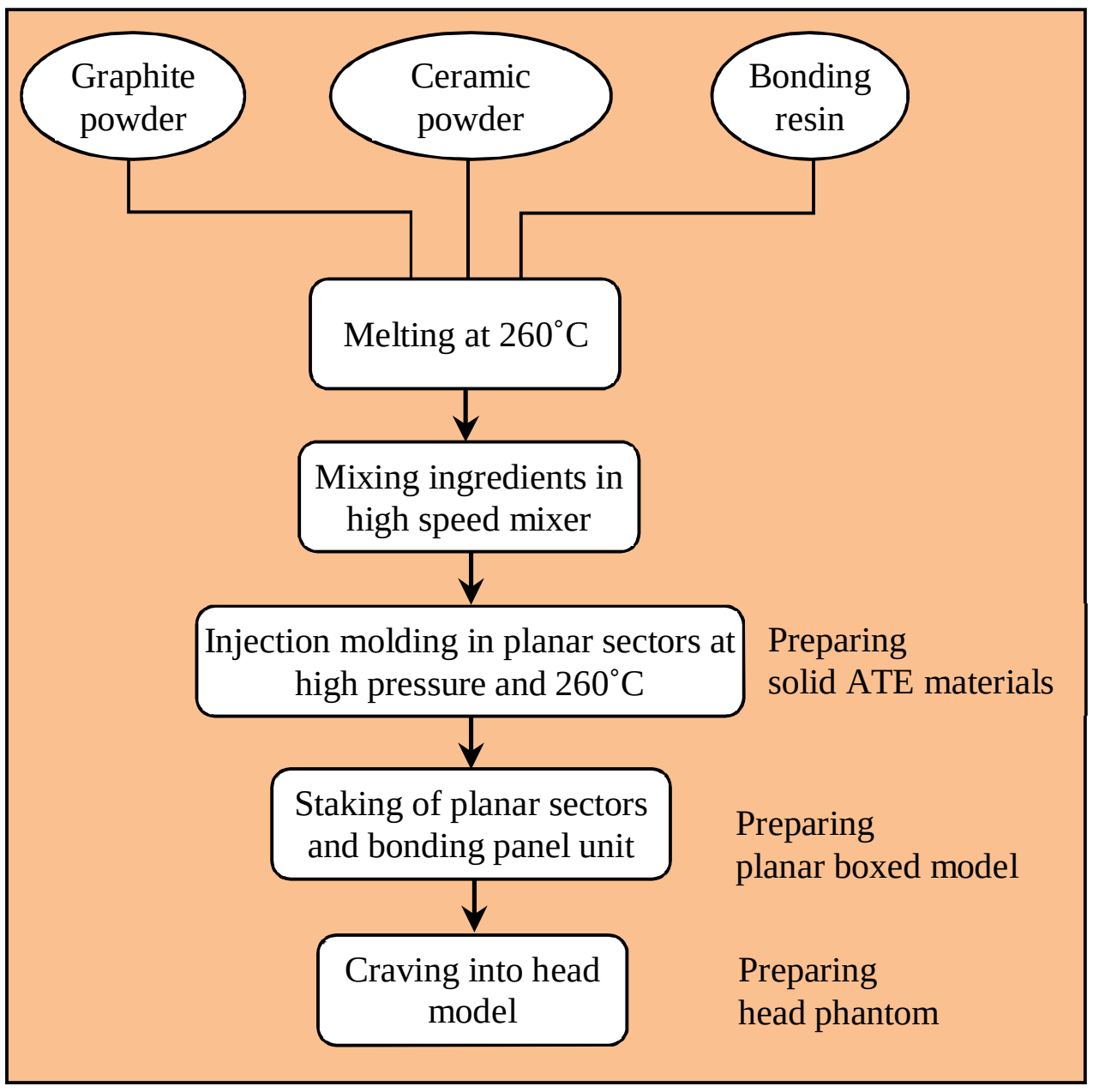

Fig. 11

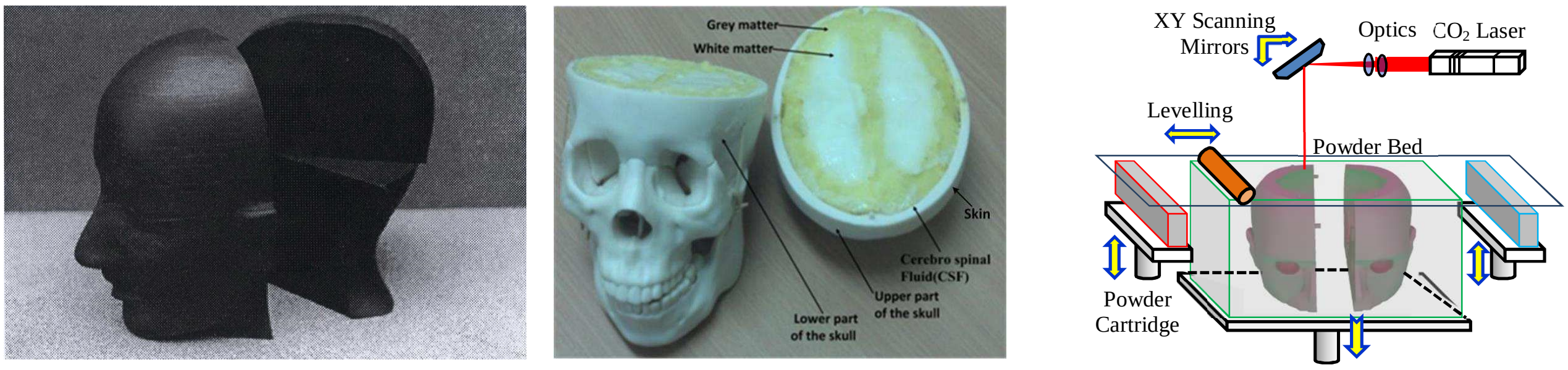

(a) (b) (c)

Fig. 12

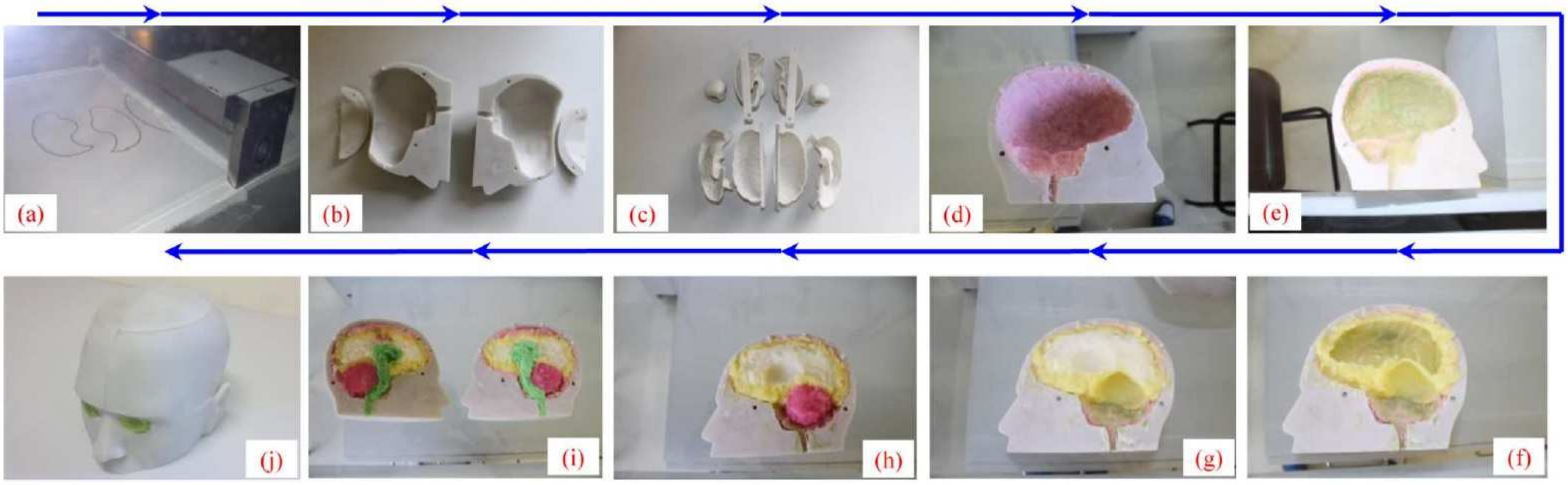

Fig 13

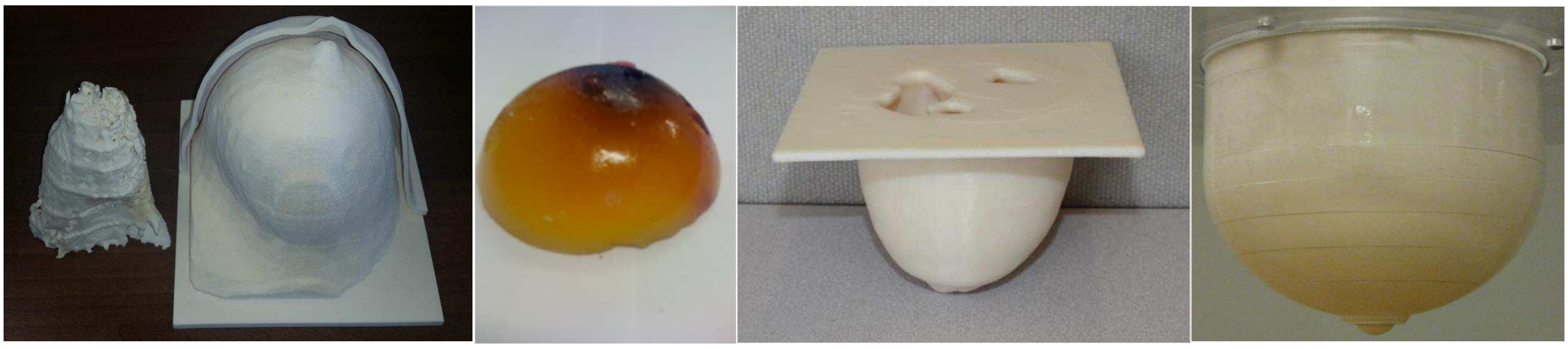

(a) (b) (c) (d)

Fig 14

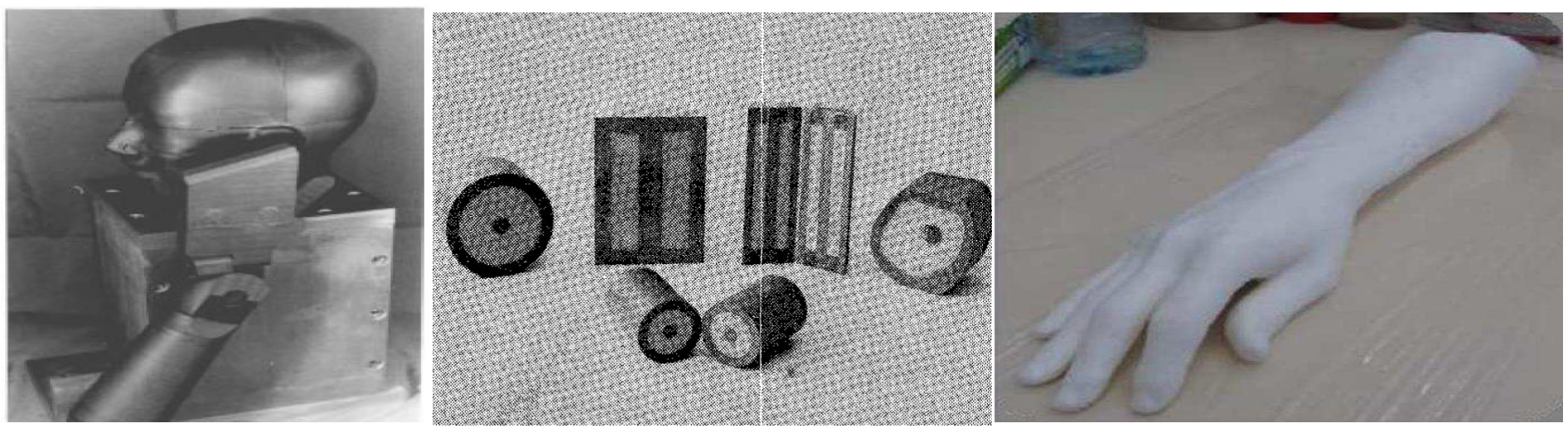

(a) (b) (c)

Fig 15

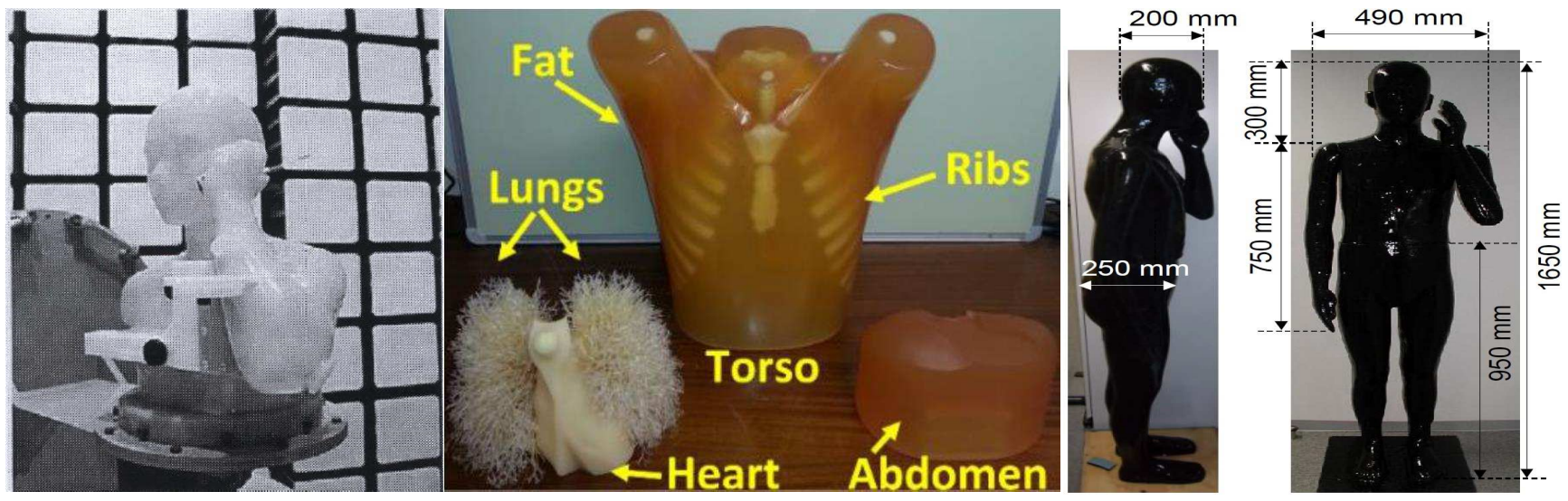

(a) (b) (c)

Fig 16

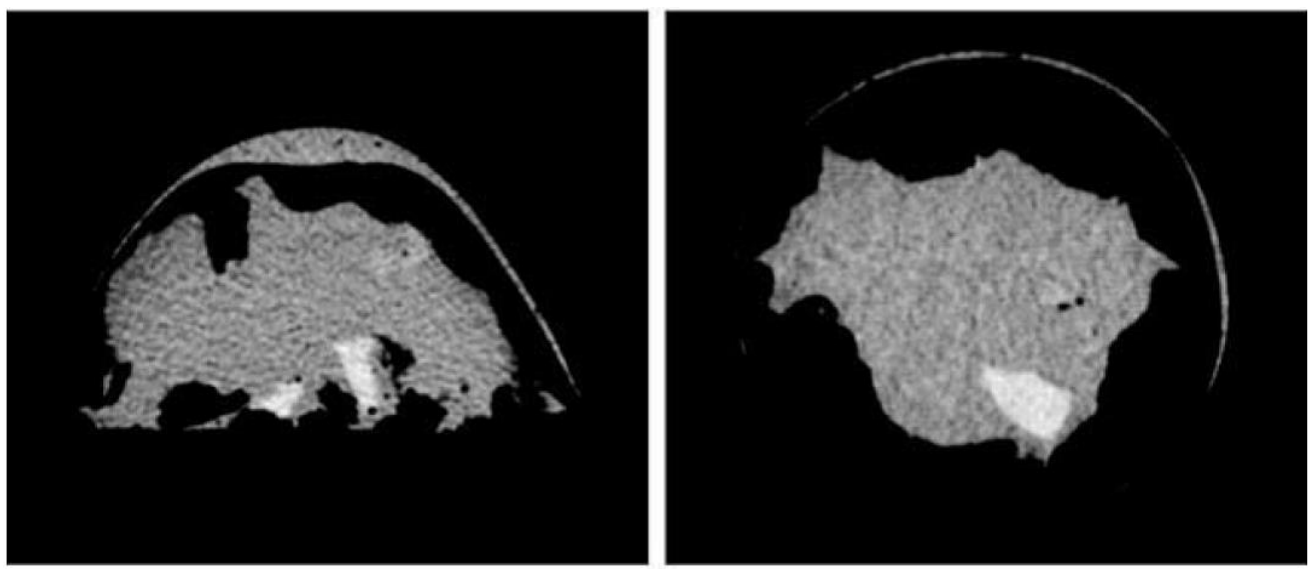

(a) (b)

**Table I: Electrical properties of different main tissues of the human body across the band 500 MHz to 10 GHz.**

| Tissues | Muscle | Blood | Fat | Nerve | Grey Matter | White Matter | Lung Inflated | Dura | CSF | Bone | Heart | Tongue | Vitreous Humour | Stomach | Dry Skin |
|---|---|---|---|---|---|---|---|---|---|---|---|---|---|---|---|
| $\varepsilon_r$ | 56.5-42.8 | 63.3-45.1 | 5.54-4.6 | 34.5-23.8 | 56-38.1 | 41-28.4 | 23.2-16.2 | 46-33 | 70.1-52.4 | 5.6-4.6 | 64-42 | 57-41.5 | 69-58 | 66.7-49 | 45-31.3 |
| $\sigma$(S/m) | 0.8-10.6 | 1.38-13.1 | 0.04-0.6 | 0.5-6.03 | 0.8-10.3 | 0.47-7.3 | 0.4-4.21 | 0.9-8.6 | 2.3-15.4 | 0.03-0.6 | 1.02-11.8 | 0.8-11.1 | 1.54-15.1 | 1.04-13.3 | 0.73-8 |

**Table II: Summary of the reported head phantoms**

| Ref. | Freq. | T or P | Type of tissue/ phantom | Included tissues | Head phantom structure /shape | Application |
|---|---|---|---|---|---|---|
| [30] | 7 MHz | P | G, HO | Brain | ANM | Medical sensors |
| [37] | 13.6, 27, 40.7, 2450 MHz | T | G, HO | Muscle, fat, skin, brain | - | Hyperthermia |
| [29] | 400 MHz | P | L | Grey matter | ANM | Medical implants |
| [72] | 402 MHz | P | L, HO | Skin | ANM, Cubic | Implantable devices |
| [55] | 0.9 GHz | P | L, HO | Tissue-equivalent liquid | ANM | SAR |
| [56] | 0.9 GHz | P | L, ML | Bone, skin, muscle, brain (average of grey and white matter), eye | ANM /realistic | SAR |
| [73] | 0.9 GHz | P | L, HO | Human brain equivalent liquid | Stylized | Antenna performance analysis |
| [2], [14]-[19] | 900 MHz, 1.5 GHz | P | S, HO | Average head equivalent | ANM | SAR |
| [57] | 0.9, 1.75, 1.95 GHz | P | L, HO | Head tissue equivalent | ANM | SAR |
| [58] | 0.9, 1.8 GHz | P | L, HO | Head simulation liquid | ANM | SAR |
| [39] | 0.9, 1.8, 1.9, 2 GHz | P | L, HO | Human tissue equivalent | ANM | SAR |
| [40] | 2.4 GHz | P | L, HO | Human brain simulant | ANM | Hyperthermia |
| [67] | 2.4 GHz | P | S, HE | Skin, cortical bone, grey and white matter, muscle | Stylized | Wearable systems |
| [68] | 2.45 GHz | P | S, HE | Skin, cortical bone, grey matter and cerebellum, muscle | Box shaped | Wearable systems |
| [60] | 2.45 GHz | P | G, HE | Bone/fat, muscle, grey matter, white matter, blood | CT scan based realistic shaped | Microwave tomography |
| [41] | 3.4 GHz | T | SS, HE | Neonate brain tissue | Modelled | Microwave radiometry |
| [74] | 1M-10 GHz | P | S, ML | Grey matter, fat, muscle | Layered, spherical | SAR measurement |
| [44] | 0.1 - 1 GHz | T | L, HO | Bone (cast, liquid), brain, and muscle | - | Hyperthermia |
| [17], [62] | 0.2-3 GHz | P | SS, ML | Brain, skull | Cubic, spheric | SAR |
| [61] | 0.3-2.5 GHz | P | G, HO | Muscle, brain | Cubic | Hyperthermia |
| [23], [49] | 0.5-4 GHz | P | SS, HE | Head exterior, grey matter, white matter, dura, CSF, eye, cerebellum, spinal cord, blood | MRI devolved realistic | Head imaging, SAR |
| [69] | 835-925 MHz | T | S, HO | Muscle, brain, skull | - | Electromagnetic dosimetry |
| [22], [63], [64], [70] | 1-4 GHz | P | SS, HE | Hair (normal/dyed), scalp, skull, CSF, grey (dead/alive), white (dead/alive), blood | ANM | Microwave brain imaging |
| [59] | 1.1-1.6 GHz | P | L, HO | Scalp, skull, brain | ANM | Radiometric monitoring |
| [71] | 1-4 GHz | P | SS, HO | Skull, brain | ANM | Head imaging |
| [65], [8] | 2-5 GHz | P | SS, ML | Skin, bone, dura, grey matter, and white matter | Stylized | Implantable electronics |
| [66] | 3-6 GHz | T | SS, HO | Head tissue equivalent | Box shaped | SAR |

* T = Tissue, P = Phantom, HO = Homogeneous, HE = Heterogeneous, ML = Multilayered; L = Liquid, G = Gel, SS = Semi-solid, S = Solid; ANM = Anthropomorphic;

**Table III: Summary of the reported breast phantoms**

| Ref. | Freq. | T or P | Type of tissue/ phantom | Included tissues | Breast phantom structure /shape | Application |
|---|---|---|---|---|---|---|
| [78] | 2.45 GHz | P | L, HO, | Breast tissue equivalent, tumour | Stylized, cylindrical | Microwave breast mammography |
| [81], [82] | 0.2-6 GHz | P | SS, HE | Fat, gland, skin, tumour | ANM, hemispherical | Microwave imaging |
| [45] | 0.5-6 GHz | T | L, HO | Normal (3 types) and malignant breast tissues | - | Microwave imaging |
| [79] | 0.5-12 GHz | T | L, HO | Normal (3 types) and malignant breast tissues | - | breast cancer detection |
| [77] | 0.5-20 GHz | T | SS, HE | Muscle, wet skin, dry skin, fat, cancerous lesions | - | Microwave applications |
| [7] | 0.5-4 GHz | P | SS, HE | Skin, adipose (fat), fibroglandular tissue, tumour | MRI-derived realistic | Microwave imaging |
| [83], [84] | 0.5-6 GHz | P | SS, HE | Fat, transitional, fibroglandular, and skin tissues | Stylized, cylindrical | Microwave imaging |
| [85] | 0.5-8 GHz | P | SS, HE | Skin, fat, fibroglandular and malignant tumor tissues | Stylized, cylindrical | Microwave breast imaging |
| [80] | 0.5-3.5 GHz | P | S, HE | Adipose tissue, fibroglandular tissue | MRI-derived realistic | Microwave breast imaging |
| [48] | 1-6 GHz | P. | SS, HE | Skin, fat, fibroglandular, tumour | Realistic | Microwave breast imaging |
| [86] | 1-11 GHz | P | L, ML | Skin, fatty breast tissue, malignant tissue | Stylized | Microwave imaging |
| [76] | 1-13 GHz | P | G, HE | Skin, gland, fat, tumour | Realistic | Radar-based microwave imaging |
| [90], [91] | 2-12 GHz | P | S, HO | Average breast tissues | ANM | Ultra-wideband imaging |
| [11], [77], [89] | 3-10 GHz | P | SS, HE | Skin, dense tissue, tumour, normal breast tissue | ANM | Breast cancer detection |
| [75], [92] | 3-11 GHz | P | SS, HE | Fat, tumour, glandular tissues | Hemispherical | Ultra wideband imaging |

**Table IV: Summary of the reported limb phantoms**

| Ref. | Freq. | T or P | Type of tissue/ phantom | Included tissues | Limb phantom structure /shape | Application |
|---|---|---|---|---|---|---|
| [38] | 80–500 MHz | P | SS, HE | Fat, muscle, tumor and marrow filled bone | ANM | RF heating and MRI thermal monitoring verification |
| [1], [93] | 0.2-2.5 GHz | P | SS, HE | Fat, bone, muscle | ANM | Microwave diathermy application |
| [94] | 0.6-6 GHz | P | S, HO | Average properties of all hand tissues (blood, muscle, tendon, bone, fat and skin) | ANM | SAR measurement |
| [19] | 0.9 GHz | P | S, HO | Average human tissues | ANM | Study of SAR, hyperthermia and antennas close to human body |
| [95], [96] | 55-65 GHz | P | SS, HO | Skin-equivalent | ANM | Body area networks |
| [35] | 57-64 GHz | P | SS, HO | Skin-equivalent | Realistic | Wireless body area networks |

**Table V: Summary of the reported torso phantoms**

| Ref. | Freq. | T or P | Type of tissue/ phantom | Included tissues | Torso phantom structure /shape | Application |
|---|---|---|---|---|---|---|
| [47] | 300, 433, 750, 915, 2,450 MHz | T | G, HO | Muscle | - | RF hyperthermia |
| [73] | 2,450 MHz | T | G, HO | Muscle, fat, skin | - | Electromagnetic hyperthermia |
| [97] | 225 MHz | T | SS, HO | Liver tumor tissue | - | hyperthermia safety |
| [99] | 451 MHz | T | SS, HO | Fat, muscle | - | Hyperthermia |
| [107] | 402 MHz, 2.4 GHz | P | G, HO | Skin | Assumed | glucose-monitoring |
| [19] | 868 MHz | P | G, ML | Skin, fat, and muscle | Modelled, multilayered | Implanted RFIDs |
| [108], [33] | 2.45 GHz | P | L, HO | Muscle | Box shaped | wearable antenna measurements |
| [109] | 2.45 GHz | T | L, HO | Muscle, fat | - | Electric-field pattern mapping |
| [110] | 8.5, 10 GHz | T | G, HE | Muscle, bone, fat | - | Dosimetry studies |
| [111] | below 30 MHz | T | G | Muscle | - | Body area network |
| [100] | 5-40 MHz | T | SS | Various tissues | - | Hyperthermia, SAR |
| [101] | 10-50 MHz | P | SS | Muscle-equivalent | Assumed | Hyperthermia |
| [44] | 0.1-1 GHz | T | L, S, HO | Bone (cast, liquid), lung, brain, and muscle | - | Electromagnetic dosimetry |
| [42] | 0.3-20 GHz | P | SS, ML | Wet skin, fat, blood, muscle | Modeled | Monitoring of blood glucose |
| [112] | 0.9-3 GHz | T | L, HO | Average properties of torso | - | SAR |
| [102] | 0.9-10 GHz | T | SS, HO | 2/3-muscle | Stylized | UWB communications |
| [98] | 1.1-1.65 GHz | P | L, HE | Fat, muscle, kidney and urine | Stylized | Microwave radiometry |
| [103] | 1-10 GHz | T | S, HO | Skin, bone, fat | - | Biomedical applications |
| [66] | 3-6 GHz | T | SS, HO | Body, head | - | SAR measurement |

**Table VI: Features and functions of the main ingredients reported for phantom fabrication procedures**

| Name | Features | Functions/Benefits |
|---|---|---|
| Water | Main solvent or constituent of ATE material | Primarily contributes to frequency dispersive high $\varepsilon_r$ |
| Sodium chloride (NaCl) | Crystallized salt | Increases imaginary part of $\varepsilon_r$ and ionic $\sigma$, while slightly decreases real part of $\varepsilon_r$; however, NaCl has no influence above 25 GHz. |
| Sucrose (Sugar) | White, crystalline, odorless powder | Used to significantly tune down relative $\varepsilon_r$, while also slightly increases $\sigma$ |
| TWEEN | Nonionic, viscous liquid detergent material | Emulsifying agent for stable oil-in-water emulsions |
| Triton X-100 | Viscous nonionic surfactant fluid. | Used as emulsifier in ATE liquids; reduce $\varepsilon_r$ |
| TX-150 (super stuff) | Water-soluble powder | Thickener, and to increase viscosity and homogeneity |
| Sodium azide ($NaN_3$) | White powder, however toxic | Generally, used as a preservative; also to linearly increase $\sigma$ in water solution |
| Ceramic powder (Ba, Ca)(Ti, Sn)O) | Can be pulverized into powders with particle size of 30 µm. | To increase $\varepsilon_r$ of the ATE materials |
| Graphite (carbon) powder | A semiconductive material that presents electrical properties when covered with bonding resin (e.g. Poly Vinylidene Fluoride (PVDF)) | To increase $\sigma$ of the ATE materials |
| Hydroxyetheylcellulose (HEC) | A nonionic water-soluble polymerizing agent (gelling agent) | Increases viscosity of water based compounds |
| Agar, Gelatin | Both are gelatinous substances; agar is yellowish white powder and gelatin is colorless | As gelling agent to hold artificial tissues to its shape and tune the dielectric properties |
| Oil (Paraffin, grape seed oil, Kerosene, safflower oil, etc) | A hydrophobic viscous liquid | To fabricate low-water-content ATE materials |
| Carbon fibre, aluminium powder | $\mu m$ thick fibers of carbon, powdered aluminium | Enables refinement of the process in different semi-solid ATE materials |
| Silicone (not to be confused with Silicon (Si)) | A heat-resistant, rubber-like synthetic polymer | Provides a matrix to hold the active ingredient; carries required amount of carbon in solid ATE materials and cure to the right consistency; silicone emulsion is utilized to control $\varepsilon_r$. |
| Acrylamide ($C_3H_5NO$) | Reacts with water to form polyacrylamide | Used in polymerization reaction, lowers $\varepsilon_r$ and slightly increases $\sigma$, increases mechanical strength, slightly impairs transparency. |
| Bactericide, preservative | Several substances that resists the growth of bacteria | To prevent decomposition of ATE materials by microbial growth to enhance shelf lifetime |
| TX151 | modified polysaccharide with propyl-paraben preservative | As a gelling agent, also functions to prevent thermal convection currents, mimics tissue texture and increases stickiness |
| Polyvinyl chloride (PVC) | White polymer powder | For exponentially decreasing $\sigma$ and linearly decreasing $\varepsilon_r$ |